\documentclass{ws-procs975x65}
\usepackage{epstopdf}
\begin{document}



\title{3-D GRMHD and GRPIC SIMULATIONS OF DISK-JET COUPLING AND EMISSION}

\author{K.-I. NISHIKAWA and Y. MIZUNO}

\address{National Space Science and Technology Center,
320 Sparkman Drive, Huntsville, AL 35805 USA\\
\email{Ken-Ichi.Nishikawa@msfc.nasa.gov}}

\author{M. WATSON}

\address{Fisk University, Department of Physics,
1000 17th Ave North, 
Nashville, TN 37208 USA\\
\email{mwatson@fisk.edu}}

\author{P. HARDEE}

\address{The University of Alabama, Tuscaloosa, AL 35487, USA}

\author{S. FUERST}

\address{KIPAC, Stanford Linear Accelerator Center,
      2575 Sand Hill Road, MS 29, Menlo Park, CA 94025 USA}

\author{K. WU}

\address{MSSL, University College London, Holmbury St.~Mary,
Surrey RH5 6NT, UK}



\author{G. J. FISHMAN}

\address{NASA/MSFC
   320 Sparkman Drive, VP62
   Huntsville, AL 35805 USA\\}

\begin{abstract}
We investigate jet formation in black-hole systems using 3-D General
Relativistic Particle-In-Cell (GRPIC) and 3-D GRMHD simulations.
GRPIC simulations, which allow charge separations in a collisionless
plasma, do not need to invoke the frozen condition as in GRMHD
simulations. 3-D GRPIC simulations show that jets are launched from
Kerr black holes as in 3-D GRMHD simulations, but jet formation in
the two cases may not be identical. Comparative study of black hole
systems with GRPIC and GRMHD simulations with the inclusion of
radiate transfer will further clarify the mechanisms that drive the
evolution of disk-jet systems.
\end{abstract}

\bodymatter

\section{GRPIC numerical simulations and initial results}

So far the black hole system has been investigated only by GRMHD
simulations, which ignore various important kinetic effects, in
particular, charge separation due to different motions of electrons
and positrons/ions in a magnetic field
\cite{Nis05,Miz06a,Miz06b,Miz06c}.
We investigated magnetic coupling in disk-jet system with proper
treatment of kinetic physics, using our general relativistic
particle-in-cell (GRPIC) code (for details, see Watson et al.
2006\cite{Wat06}). Our objective is to identify the essential
micro-physics for jet formation.

The underlying physics of the particle motion is the contravariant
form of the Newton-Lorentz equation. This form provides the equation
for  the acceleration of the particle. The acceleration is a
function of the spacetime curvature defined by the metric and the
Lorentz force due to the electromagnetic field. The local field is
described by the Maxwell field tensor. The components of the tensor
are calculated using the contravariant general relativistic form of
Maxwell's equations. Using these three equations, the particles are
moved and the fields and currents are calculated self-consistently.
The individual kinematics of the system governs the evolution of the
simulation. The algorithm uses the relativistic  PIC method
\cite{Nis06}.
The particle motion is calculated
by integrating the equation of motion using a fourth order
Runge-Kutta. Similarly to the RPIC, the electric and magnetic field
components of Maxwell's field tensor are offset in space and time.

We present results of our 3-D simulations of jet formation using
general relativistic plasma particle dynamics in Kerr metric
(angular momentum $a=J/J_{\max} = 0.9$). The initial setting of the
simulation is as follow: a background plasma (8 electron-positron
pairs/cell), a free falling corona, and a Keplerian disk as that in
GRMHD Simulations\cite{Nis05,Miz06a,Miz06b,Miz06c}.
The black hole is located at the origin. The particle number of the
corona is $1/100$ of the disk. The Keplerian disk is located at
$r>r_{\rm D}\equiv 3r_{\rm S}$ $|\cos{\theta}| < \delta$, where
$\delta=1/8$, $r_{\rm D}$ is the disk radius and $r_{\rm S} \equiv
2GM/c^2$ is the Schwarzschild radius. In this region the particle
number is 100 times that of the corona. The orbital velocity of the
particles in the disk is $v_\phi = v_{\rm K} \equiv c\sqrt{r_{\rm
S}/(2r)}$, where $v_{\rm K}$ is the Keplerian velocity, $r =
\sqrt{-a^2/2+R^2/2 +\frac{1}{2}\sqrt{(a^2-R^2)^2+4a^2 z^2}}$, and
$R^2=x^2+y^2+z^2$\cite{Bel95}. 
There are no disk particles initially at $r<r_{\rm D}$.  The initial
magnetic field is taken to be uniform in the $z$ direction. The
magnitude of the field is $10^4$ gauss. This field component is the
contravariant $z$ component of the field.

Figure 1 shows GRPIC simulations of co-rotating jets launched from
the disk along the $y-z$ plane at $x=0$. The angular momentum of
particles shearing towards the black hole is transferred to the jet,
causing it to spiral. The particles are ejected away from the black
hole, which leads to formation of bipolar jets. Figure 1 also shows
a 3-D view of particle velocities. The most remarkable phenomenon is
that the jet is actually composed of streams of particles with
different velocities. Such streams are kinematically unstable,
easily giving rise to non-linear and collective behaviors, which
leads to bunching, clumping and shock formation. Also noticeable is
the development of spiral structure (color). We note that the jet
structure depends upon the plasma particle species assumed in the
simulation. At this time there is a need to quantify the parameters
required for jet formation. Our study of particle ejection has
demonstrated that jet material consists of particle pairs, i.e. not
restricted to baryons as assumed in many theoretical models. The
particles in the jet maintain some angular momentum and continue to
spiral around the central axis.

\begin{figure}[htbp] 
   \centering
   {\includegraphics[angle=-90,width=.35\textwidth]{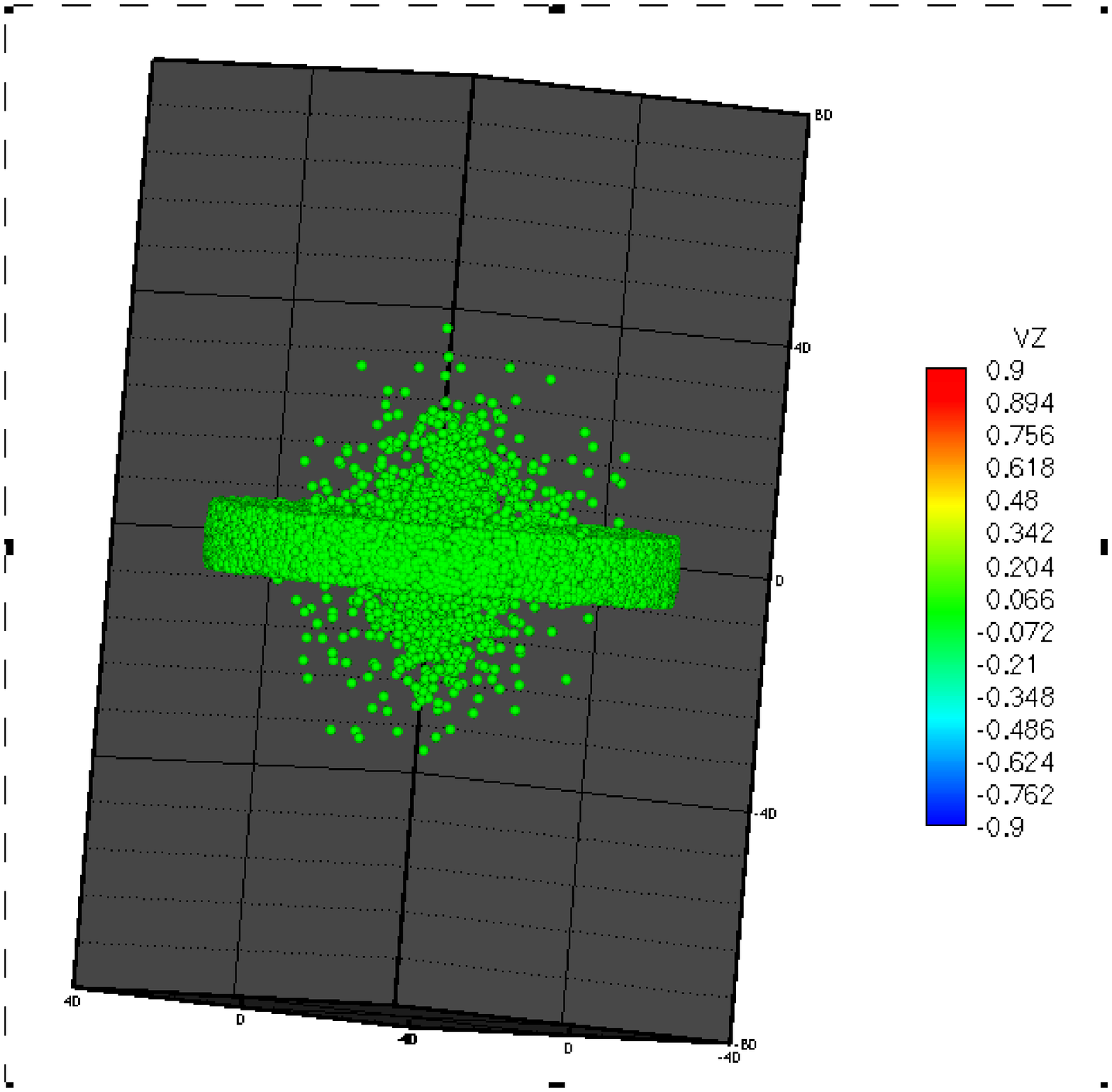}}
   \hspace{.2in}
   {\includegraphics[angle=-90,width=.35\textwidth]{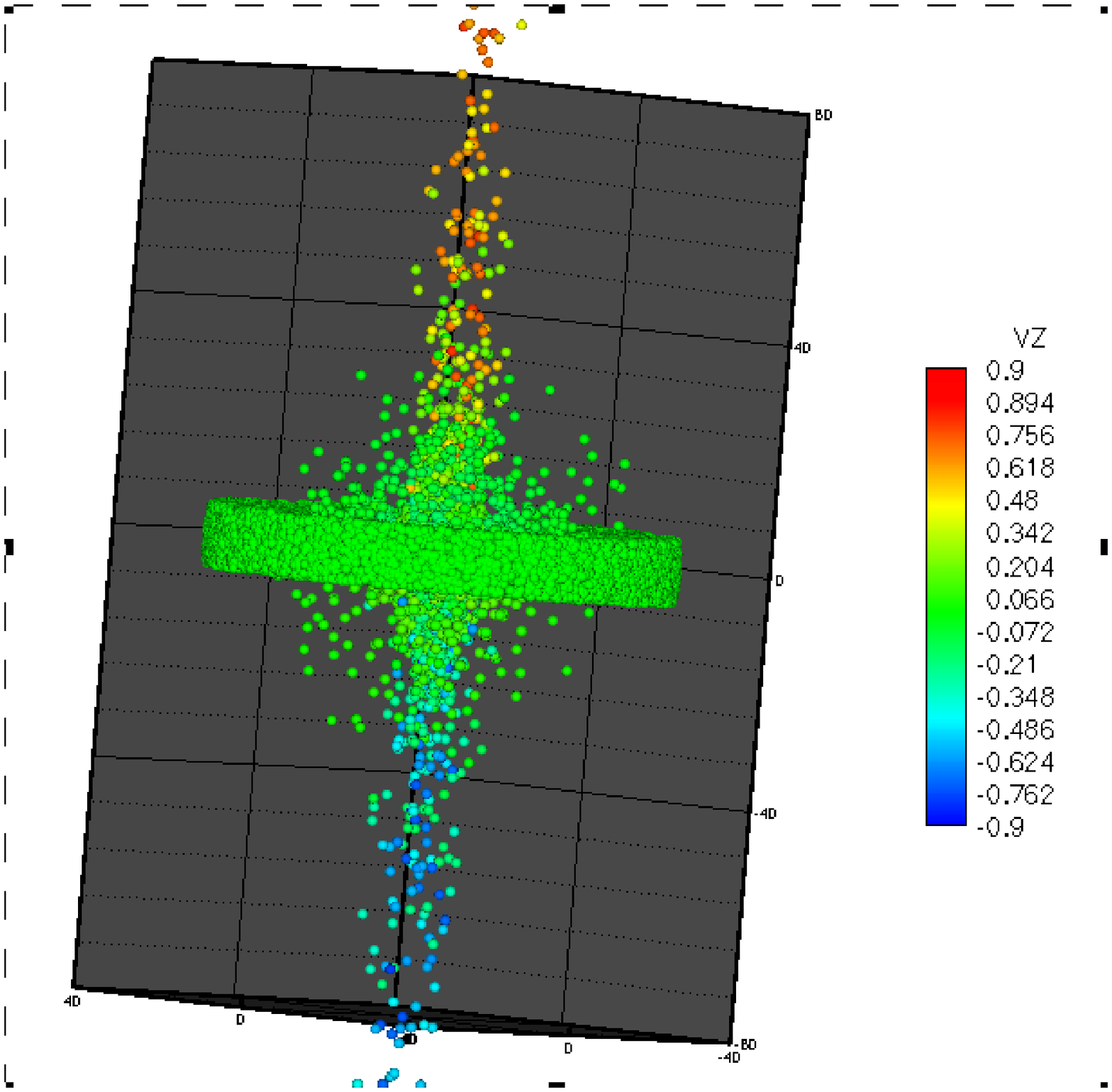}}
\caption{The 3D views of disk and corona particles are shown at
$t/\tau_{\rm S} = 0$ and 1134 ($\tau_{\rm S} = r_{\rm S}/c$).
Particle pairs are
   moving through the jet at different velocities. The jet has a structure
   which forms spirals around the $z$ (central)  axis.}
 \label{jetvel}
\end{figure}

The simulations do not show significant magnetic field deformation.
This may be a consequence of the number and the strength of
particles prescribed in the simulation. More particles may cause a
stronger disk magnetic field and hence a stronger electric field,
due to the particles falling towards the horizon. Another possible
explanation for the lack of significant field deformation is that
the frozen-in condition is relaxed. In GRMHD simulations, as the
frozen-in condition $\vec{E} = -\vec{v} \times \vec{B}$ is used, the
magnetic fields are dragged by the fluid motion. The relative
significance of each of these effects is crucial for proper
understanding of jet formation and needs to be quantified.

The inclusion of radiation transfer (Fuerst et al.2006\cite{Fue06})
is essential to make observational predictions and hence to verify
the findings in the GRPIC and GRMHD simulations.

\section*{Acknowledgments}

K.~N. acknowledges partial support by National Science Foundation
award AST-0506719 and the National Aeronautic and Space
Administration awards NNG05GK73G and HST-AR-10966.01-A to the
University of Alabama in Huntsville. M.~W. acknowledges partial
support by a NASA Faculty Fellowship Program at National Space and
Technology Center.

\end{document}